\documentclass{aa}
\usepackage{epsf}
\usepackage{graphicx}

\newcommand{\lesssim}{\mathrel{\hbox{\rlap{\hbox{\lower4pt\hbox{$\sim$}}}\hbox{$
<$}}}}
\newcommand{\gtrsim}{\mathrel{\hbox{\rlap{\hbox{\lower4pt\hbox{$\sim$}}}\hbox{$>
$}}}}
\newcommand{\sigmat}{\sigma_\mathrm{T}}
\newcommand{\Nh}{N_\mathrm{H}}

\newcommand{\flu}{S_\mathrm{X}}
\newcommand{\flux}{F_\mathrm{X}}

\begin{document}

\title{Observing scattered X-ray radiation from gamma-ray bursts: a way
to measure their collimation angles}

\author{S.Yu. Sazonov$^{1,2}$ and R.A. Sunyaev$^{1,2}$} 

\offprints{sazonov@mpa-garching.mpg.de}

\institute{$^1$Max-Planck-Institut f\"ur Astrophysik,
              Karl-Schwarzschild-Str. 1, 
              85740 Garching bei M\"unchen, Germany\\
              $^2$Space Research Institute, Russian Academy of Sciences,
              Profsoyuznaya 84/32, 117997 Moscow, Russia\\}

\date{Received 1 July 2002; accepted 22 November 2002}

\titlerunning{Scattered X-ray emission from gamma-ray bursts}

\authorrunning{Sazonov \& Sunyaev}

\maketitle

\begin{abstract}
There are observational facts and theoretical arguments for an origin
of gamma-ray bursts (GRBs) in molecular clouds in distant galaxies. If
this is true,  one could detect a significant flux of GRB prompt and
early afterglow X-ray radiation scattered into our line of sight by
the molecular and atomic matter located within tens of parsecs of the
GRB site long after the afterglow has faded away. The scattered flux
directly measures the typical density of the GRB ambient medium. Furthemore, if
the primary emission is beamed, the scattered X-ray flux will be
slowly decreasing for several months to years before falling off 
rapidly. Therefore, it should be possible to estimate the collimation 
angle of a burst from the light curve of its X-ray echo and a measured
value of the line-of-sight absorption column depth. It is shown that
detection of such an echo is for the brightest GRBs just within the
reach of the Chandra and XMM-Newton observatories.


\end{abstract}

%%%%%%%%%%%%%%%%%%%%%%%%%%%%%%%%%%%%%%%%%%%%%%%%%%%%%%%%%%%%%%%%%%%
\section{Introduction}
\label{intro}
%%%%%%%%%%%%%%%%%%%%%%%%%%%%%%%%%%%%%%%%%%%%%%%%%%%%%%%%%%%%%%%%%%%

One of the fundamental questions relating to the gamma-ray burst (GRB)
phenomenon is how much energy is released during a GRB. Given a 
measured burst fluence, the inferred energy release is proportional
to $\theta_0^2/2$, where $\theta_0$ is the opening angle of the
collimated relativistic fireball (Piran 1999) that is probably small
($\theta_0\ll 1$). Apart from constraining the burst energetics,
$\theta_0$ also determines the event rate of GRBs in the universe
(which is proportional to $\theta_0^{-2})$. Thus, the beaming angle is
a very important quantity in GRB research.

There have been attempts to estimate $\theta_0$ via observing a break
in GRB afterglow light curves and interpreting this break as due to a
slowing down of the jet to $\Gamma<\theta_0^{-1}$ (here $\Gamma$ is the
bulk Lorentz factor of the jet). Summarizing the information on jet break
times in more than a dozen of GRBs Frail et al. (2001) have inferred
values for $\theta_0$ ranging from $1^\circ$ to more than
$25^\circ$. These authors came to the conclusion that the gamma-ray 
energy release is narrowly clustered around $5\times
10^{50}$~ergs. However, since their derivation was based on a 
particular interpretation of the observed breaks and that other
explanations for such breaks exist (as mentioned by Frail et
al. 2001), these measurements of $\theta_0$ should necessarily be considered
tentative for the time being.

Another fundamental issue is the environment in which GRBs
occur. High gas column densities ($\Nh\sim 10^{22}$~cm$^{-2}$) toward
GRB locations have been inferred from a spectral analysis of a sample of 
X-ray afterglows observed with BeppoSAX (Owens et al. 1998; Galama et al. 2001;
Reichart \& Price 2002). Such values of $\Nh$ are typical of Galactic
giant molecular clouds and therefore a strong indication that GRBs
occur in star forming regions of their host galaxies, which in turn
argues for collapse of massive stars as their sources (Woosley 1993;
Paczy\'{n}ski 1998). 

In this paper we propose an observational test that enables
direct determination (with some reservations) of both the collimation
angle of a GRB and the typical density of the medium surrounding it on
scales of pcs to tens of pcs typical of molecular clouds and
complexes. This method consists of observing the location of a bright
GRB with a powerful X-ray telescope several times during the first
months to years after the burst. We predict that in such observations
a weak X-ray flux may be detected, which will be radiation from the
burst reaching us, delayed by scattering from the ambient medium. 

The idea at the basis of the proposed method -- to search for
scattered GRB radiation -- is not new. The theme of scattered X-ray
emission was first introduced by Dermer, Hurley \& Hartmann (1991) in
the context of the then popular model of the galactic stellar binary origin of
GRBs. Madau et al. (2000) considered Compton echoes from gamma-ray
bursts arising in the circumburst environment within a fraction of a
pc of the GRB site. These authors focused on the case where the
primary burst emission is collimated away from us. Ramirez-Ruiz et
al. (2001) furtherelaborated this scenario. Esin \& Blandford
(2000) studied the scattering of GRB early optical emission on the
surrounding dust, while M\'{e}sz\'{a}ros \& Gruzinov (2000) considered
a similar phenomenon -- small-angle scattering of X-rays by dust. 

Ghisellini et al. (1999) and B\"{o}ttcher et al. (1999) have computed the time 
dependence of fluorescence emission in the iron $K\alpha$ line
resulting from the interaction of the GRB radiation with the surrounding
matter under the assumption that the burst emission is
isotropic. Detection of such a fluorescent line would make it possible
to determine the redshifts of GRBs with invisible optical
afterglows. However, the predicted flux in the line is apparently too low to be
detected with the current generation of X-ray telescopes unless the density of
interstellar material at the GRB site is very high ($n\gtrsim
10^4$~cm$^{-3}$). We note here (see \S\ref{effect}) that the scattered
X-ray flux in the relevant 0.3--5~keV band will typically be an order
of magnitude higher than the flux of fluorescence emission in the same
band. As we shall see, this enhancement factor proves crucial for the
problem at hand, as it makes the emergent signal detectable with the
present-day X-ray telescopes provided that GRBs indeed originate
within molecular clouds.

%%%%%%%%%%%%%%%%%%%%%%%%%%%%%%%%%%%%%%%%%%%%%%%%%%%%%%%%%%%%%%%%%%%
\section{Delayed scattered X-ray emission from GRB}
\label{effect}
%%%%%%%%%%%%%%%%%%%%%%%%%%%%%%%%%%%%%%%%%%%%%%%%%%%%%%%%%%%%%%%%%%%

\begin{figure}
\centering
\includegraphics[width=\columnwidth]{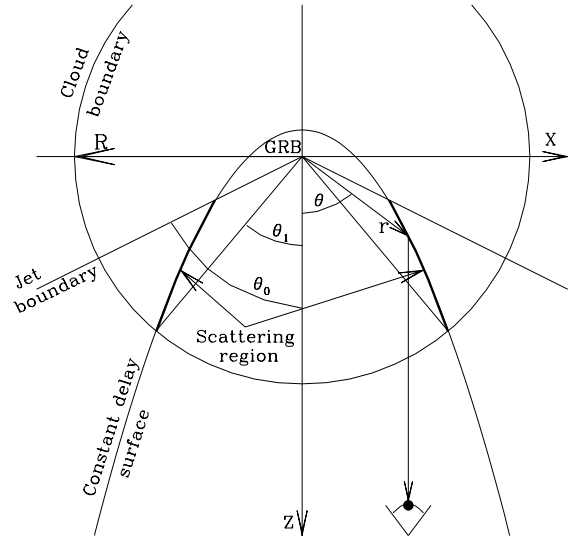}
\caption{Schematic diagram (cross section) of the model in which a
collimated GRB occurs at the center of a molecular cloud and the observer sees
GRB X-ray radiation scattered by the gas. The scattering sites 
contributing to the detected flux at a given time lie on a
constant-delay paraboloid between the jet and cloud boundaries.
}
\label{sketch}
\end{figure}

We base our treatment on a simple model depicted in
Fig.~\ref{sketch}. A GRB occurs at the center of a spherical cloud of
gas of constant density $n$ (equivalent number density of hydrogen
atoms; the gas may be a mixture of atoms and molecules) and radius
$R$. The host galaxy is located at a redshift $z$. The blastwave
expands into a cone with an opening angle $\theta_0$, so that only
observers located within this cone can directly receive X-ray and
gamma-ray radiation from the burst. Note that since the bulk
Lorentz factors of GRB fireballs are believed to be very large
($\Gamma\sim 10^2$--$10^3$), observers in other directions will detect
very low fluxes of radiation and at energies below X-rays. In our
model, we are located exactly on the symmetry axis of the jet. Another
parameter of the model is the GRB fluence, $\flu$, in some specified X-ray band
$[E_1, E_2]$. 
 
The scattered X-ray emission is observed from Earth at a time $t$ after the
GRB. The scattering surface of constant delay will be a paraboloid with
its focus at the GRB site and its axis along our line of sight,
\begin{equation}
r=\frac{ct}{(1+z)(1-\cos\theta)},
\label{front}
\end{equation}
(Blandford \& Rees 1972), where $\theta$ is the scattering angle and
$c$ is the speed of light. The factor $(1+z)^{-1}$ takes into account
cosmological time dilation. As is detailed below (in \S\ref{detect}), the
fluence of the early X-ray afterglow ($t\lesssim 10^4$~s) may for some
GRBs constitute a significant fraction of the X-ray fluence of
the burst itself. The usual interpretation for such early afterglow
emission is that it arises from the interaction of the jet with the external
medium at a distance of 0.01--0.1~pc from the site of the burst. 
Therefore, equation (\ref{front}) will remain an appropriate
approximation when considering the echo associated with the early
afterglow if $R\gtrsim 1$~pc, which we assume to be the case. 

Only positions $(r<R, \theta<\theta_0$) will contribute to the
scattered X-ray flux. The corresponding range of scattering angles is
$\theta_1<\theta<\theta_0$, where the critical angle (see Fig.~\ref{sketch})
\begin{equation}
\theta_1=\arccos\left(1-\frac{ct}{R(1+z)}\right).
\end{equation}

The scattered X-ray flux in $[E_1, E_2]$ is then
\begin{equation}
\flux=\flu n\int_{\theta_1}^{\theta_0}
\frac{d\sigma}{d\Omega}\frac{dr}{dt}\,d\Omega,
\label{flux_1}
\end{equation}
where $d\Omega=2\pi d\cos\theta$ and
$dr/dt=c(1+z)^{-1}(1-\cos\theta)^{-1}$. The differential cross section
for scattering, which includes coherent 
(Rayleigh) and incoherent (Raman and Compton) scattering, is given by
\begin{equation}
\frac{d\sigma}{d\Omega}=A(\theta)\frac{3\sigmat}{16\pi}(1+\cos^2\theta),
\end{equation}
where $\sigmat$ is the Thomson cross section. The coefficient
$A(\theta)$ allows for the fact that Rayleigh scattering of X-rays
through small angles on molecules of hydrogen and atoms of 
heavier elements, most importantly helium, is more efficient (calculated
per electron) than scattering on atomic
hydrogen. $A(\theta)$ takes values between 1 and 2 for $\theta\rightarrow 0$,
depending on the fraction of molecular hydrogen and helium in the interstellar
medium, and is close to 1 for large scattering angles (e.g. Sunyaev \&
Churazov 1996). As we are primarily interested in 
situations where the jet opening angle is small ($\lesssim 30^\circ$)
and the scattering medium is a molecular cloud, we shall adopt $A(\theta)\sim
1.5$ for our estimates below. 

\begin{figure}
\centering
\includegraphics[width=\columnwidth]{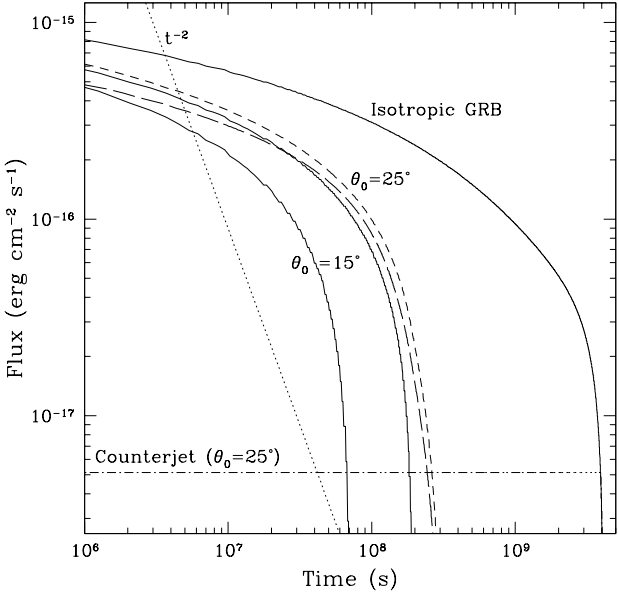}
\caption{Scattered X-ray flux as a function of elapsed time after a
GRB at $z=1$ with $\flu=10^{-5}$~erg~cm$^{-2}$. The different solid
lines correspond to a GRB origin in the center of a uniform molecular
cloud with $R=10$~pc and $n=10^3$~cm$^{-3}$, and different burst
collimation angles: $\theta_0=\pi$, $25^\circ$ and $15^\circ$. Also
shown are the light curves for a GRB with $\theta_0=25^\circ$ located at
$x=0.5 R$, $z=-0.5 R$ (see Fig.~\ref{sketch}) in the same uniform
cloud (short-dashed line) and in a cloud of radius $R=10$~pc
with a density law $n(r)=n_0(R/r)$, $n_0=5\times 10^{2}$~cm$^{-3}$
(long-dashed line). The horizontal dash-dotted line represents the
scattered signal from the counterjet. The dotted line represents X-ray
afterglow emission with parameters similar to the afterglow of the
bright GRB~000926 observed with Chandra at a late time $t\sim 10^6$~s.
}
\label{curve}
\end{figure}

On integrating over the solid angle in equation (\ref{flux_1}) we get
\begin{eqnarray}
\flux &=& \frac{3}{8}A\flu n(1+z)^{-1}c\sigmat f(\theta_0,\theta_1)
\label{flux_2}
\\
f(\theta_0,\theta_1) &=& \frac{(\cos\theta_0-\cos\theta_1)
(\cos\theta_0+\cos\theta_1+2)}{2}
\nonumber\\
&& +2\ln\frac{1-\cos\theta_0}{1-\cos\theta_1}.
\label{f_1}
\end{eqnarray}
The function $f(\theta_0,\theta_1)$ carries information on the
time dependence of the scattered flux. Introducing a critical time
\begin{equation} 
t_1=\frac{R(1+z)(1-\cos\theta_0)}{c},
\label{t1_1}
\end{equation}
we may rewrite equation (\ref{f_1}) as
\begin{eqnarray}
f &=& 2\ln\frac{t_1}{t}-\frac{1-\cos\theta_0}{2}\left(1-\frac{t}{t_1}\right)
\nonumber\\
&&\times\left[4-\left(1+\frac{t}{t_1}\right)(1-\cos\theta_0)\right]
\nonumber\\
&&\rightarrow 2\ln\frac{t_1}{t}\,\,\mathrm{for}\,t_\mathrm{GRB}\ll t\ll t_1,
\label{f_2}
\end{eqnarray}
where $t_\mathrm{GRB}$ is the duration of the burst. We notice that
the scattered flux tends to infinity as $t\rightarrow 
0$, being limited only by the finite GRB duration. This reflects the
fact that at early times the dominant 
contribution to the flux comes from small scattering angles while the
scattering front sweeps rapidly through the cloud toward us:
$dr/dt\propto\theta^{-2}\propto t^{-1}$. We also see that if the
GRB emission is collimated in a narrow cone, then the slow
(logarithmic) decline of $\flux$ will be followed by a rapid drop to
zero when the critical angle $\theta_1(t)$ approaches the jet opening
angle $\theta_0$. The scattered flux will vanish completely at $t=t_1$. 

Equations (\ref{flux_2}), (\ref{t1_1}) and (\ref{f_2}) provide a
complete description of the scattering effect. Substituting some 
values for the parameters, we get
\begin{eqnarray}
\flux &=& 5\times 10^{-16}\,\mathrm{erg}\,\mathrm{cm}^{-2}\,
\mathrm{s}^{-1}
\nonumber\\
&&\times\frac{A}{1.5}\frac{\flu}{10^{-5}\,\mathrm{erg}\,
\mathrm{cm}^{-2}}\frac{n}{10^3\,\mathrm{cm}^{-3}}\frac{1}{1+z}
\frac{f(t)}{4}
\label{flux}
\end{eqnarray}
for
\begin{equation}
t<t_1=10^9\,\mathrm{s} \frac{R}{10\,\mathrm{pc}}(1+z)(1-\cos\theta_0),
\label{t1}
\end{equation}
and $\flux=0$ for $t>t_1$. Note that $f\sim 4$ corresponds to $t\sim
0.1t_1$. 

Fig.~\ref{curve} shows examples of light curves of scattered GRB X-ray
emission. One of the light curves represents the echo from a possible
counterjet expanding away from us; the corresponding flux is very 
weak. Also plotted is the extrapolated light curve of
the X-ray afterglow of the bright GRB~000926. This afterglow was observed
with Chandra two weeks ($t\sim 10^6$~s) after the GRB, when the flux
was still decaying as a power-law with an estimated index $\alpha\sim
-2$ (Piro et al. 2001). This is the latest performed measurement of a
GRB X-ray afterglow to our knowledge and as such it may be indicative of
how steep the decay of X-ray afterglows is at $t\gtrsim 10^6$~s. Clearly,
observations aimed at detecting the scattered radiation should be
scheduled for a not too early moment of time, so that the afterglow
signal has faded away. It appears that waiting for a few months after
the GRB should typically be enough. 

Suppose now that it is possible in an experiment to measure $\flu$,
$z$, $\flux(t)$ and $t_1$. One will then be able to make the following
estimates:
\begin{eqnarray}
\theta_0 &=& \arccos\left(1-\frac{ct_1}{R(1+z)}\right)
\nonumber\\
&\approx& 25^\circ\left(\frac{t_1}{10^8\,\mathrm{s}}\right)^{1/2}
\left(\frac{10\,\mathrm{pc}}{R}\right)^{1/2}(1+z)^{-1/2}
\label{theta0}
\end{eqnarray}
and
\begin{eqnarray}
n &=& 2\times 10^3\,\mathrm{cm}^{-3}\frac{1.5}{A}(1+z)
\nonumber\\
&&\times
\frac{10^{-5}\,\mathrm{erg}\,\mathrm{cm}^{-2}}{\flu}
\frac{\flux(t)}{10^{-15}\,\mathrm{erg}\,\mathrm{cm}^{-2}\,\mathrm{s}^{-1}}
\frac{4}{f(t)},
\label{dens}
\end{eqnarray}
with $f(t)$ being given by equation (\ref{f_2}).

It is evident that by performing one or two measurements of the scattered
flux at well separated times $t\ll t_1$, i.e. in the quasi-flat part
of the light curve, it should be possible to estimate the density of
the cloud, or at least an upper limit on it, to within a factor of
2. The second (later) measurement is desirable because it enables us to
roughly constrain the shape factor $f(t)$. To determine the
collimation angle $\theta_0$ it is necessary to have a better estimate
of $t_1$, which requires at least three flux measurements at different
times to be carried out, and also to know the radius of the cloud $R$. One can
estimate the latter quantity, which is of interest in its own right,
by measuring the absorption column density toward the GRB or its afterglow:
$R=\Nh/n$. It is of crucial importance that the dependence of
$\theta_0$ on all parameters is fairly weak -- see equation (\ref{theta0}).  

The actual environment of a GRB may differ significantly from our
simplified model. However, we expect that the shape of the light curve
of the scattered emission will not be affected dramatically by
this. To illustrate this point we have plotted in Fig.~\ref{curve}
three different computed light curves corresponding to the case of
$\theta_0=25^\circ$ and $R=10$~pc, one of which is described by our
analytic solution. We see that these curves are very similar despite
the significant differences in the setup. This has a straightforward
explanation: the signal  detected at a given time results from
integration over scattering sites, satisfying the condition
$\theta_1<\theta<\theta_0$, that are located at different distances and 
in different directions from the center of the cloud. Therefore, the
scattered flux is proportional to some weighted average,
$\langle n\rangle$, of the gas density over the scattering surface,
which only weakly responds to changes in the geometry and distribution
of gas.

Although we have so far only considered the scattering process, the GRB
radiation can also be photoabsorbed by the neutral (or 
weakly ionized) matter, giving rise to fluorescence emission, mainly in
the Fe K$\alpha$ line (Ghisellini et al. 1999; B\"{o}ttcher et
al. 1999). Given the dependence of the photoabsorption cross section
on energy, $\sigma\approx 3.5\times 10^{-20}\,\mathrm{cm}^{2}
(7.11\,\mathrm{keV}/E)^{2.8}$, and the fluorescence 
yield of 0.34 (e.g. Vainshtein \& Sunyaev 1981), assuming that the 
spectrum of primary X-ray emission is $d N_\mathrm{photon}/ 
dE\propto E^{-\gamma}$ with $\gamma$ close to 1 (as typically
observed, see e.g. Amati 2002), and assuming solar abundance of iron, we find
that fluoresence radiation contributes a fraction $\sim
0.2A^{-1}(1+\cos^2\theta)^{-1}$ to the total detectable flux at  
0.3--5~keV (which corresponds to the rest-frame 0.6--10~keV at $z=1$),
i.e. $\sim 7$\% for the most interesting angles $\theta\rightarrow 0$
and a molecular fraction $A\sim 1.5$. This estimate is valid so far as
the X-ray spectrum is not heavily absorbed, i.e. for $\Nh\lesssim$~a
few $\times 10^{22}$~cm$^{-2}$, and is not strongly dependent on
$\gamma$ and $z$. We may therefore conclude that the fluorescence
emission, although interesting, will typically be an order of
magnitude fainter than the scattered component.

%%%%%%%%%%%%%%%%%%%%%%%%%%%%%%%%%%%%%%%%%%%%%%%%%%%%%%%%%%%%%%%%%%%
\section{Detectability of the effect}
\label{detect}
%%%%%%%%%%%%%%%%%%%%%%%%%%%%%%%%%%%%%%%%%%%%%%%%%%%%%%%%%%%%%%%%%%%

A key question is of course the detectability of the X-ray
echo. Equation (\ref{flux}) indicates that the effect should be just
within the reach of the currently flying Chandra and XMM-Newton
observatories for bursts with $\flu\sim
10^{-5}\,\mathrm{erg}\,\mathrm{cm}^{-2}$ occuring in molecular 
clouds with $n\gtrsim 10^3$~cm$^{-3}$. How realistic are these values
for both parameters?

First, we know that GRBs having $\flu\sim
10^{-5}\,\mathrm{erg}\,\mathrm{cm}^{-2}$ do occur from time to time. Note
that it is the X-ray fluence, rather than the total fluence, which is 
important for us. To make an estimate we have taken information (Amati et
al. 2002) on the fluxes and spectra at $>2$~keV of twelve
bright GRBs observed with the BeppoSAX satellite, then extrapolated
the spectral fits to lower energies and calculated the fluences in the
0.3--5~keV band, where both Chandra and XMM-Newton have good
sensitivity. We find two bursts among this sample with a sufficiently
high X-ray fluence: GRB~990712 ($\flu=5\times
10^{-6}\,\mathrm{erg}\,\mathrm{cm}^{-2}$) and GRB~010222 ($\flu=7\times
10^{-6}\,\mathrm{erg}\,\mathrm{cm}^{-2}$). Other gamma-ray burst
detectors with sensitivity in the X-ray band such as the one on Ginga
(Yoshida et al. 1989), GRANAT/WATCH (Sazonov et al. 1998) and recently
HETE/FREGAT (Barraud et al. 2002) have also detected a few GRBs with $\flu\sim
10^{-5}\,\mathrm{erg}\,\mathrm{cm}^{-2}$.

Fluence values quoted in the literature usually pertain to the prompt
GRB emission. However, there are indications (Burenin et al. 1999;
Giblin et al. 1999; Tkachenko et al. 2000) that a comparable X-ray
fluence may be contained in early ($\sim 10^4$~s) GRB afterglows,
during which the energy spectrum is much softer than during the burst
proper. Without attracting spectral information it cannot be possible to
separate the contributions of the prompt and early afterglow emission
to the X-ray echo at early times ($t\ll t_1$), as the scattered X-ray
flux will be simply proportional to the total of the X-ray fluences of the
burst proper and its early afterglow. However, there are two
possibilities for the light curve of the X-ray echo. In
the case where the early afterglow emission is beamed exactly as the prompt
radiation, the conclusions of \S\ref{effect} all remain true, except that
in equation (\ref{dens}) one should understand $\flu$ as the combined
X-ray fluence of the GRB and the early afterglow. If, on the other
hand, the early afterglow is characterized by its own beaming
angle $\theta_\mathrm{a}\ne \theta_0$ the light curve of 
the X-ray echo will be a sum of two similarly shaped light curves
(like those shown in Fig.~\ref{curve}), one corresponding to
$\theta_0$ and the other to $\theta_\mathrm{a}$, with their relative
weights being proportional to the fluences of the prompt and
early afterglow X-ray emission. Therefore, it should be in principle
possible to estimate from the scattered light curve the angle
$\theta_\mathrm{a}$ in addition to $\theta_0$.

As regards the amplitude of the effect, we should also mention
that Galactic absorption will lead to a reduction of the GRB fluence
and similarly of the scattered flux in the 0.3--5~keV band. This,
however, will only affect the flux near the lower boundary of the
quoted spectral range, so the net effect is expected to be small ($\sim 10$\%).

Let us next consider the gas density. This is of course a quantity
which the proposed effect enables to constrain. We may,
however, speculate a little given the information available now. As
mentioned in \S\ref{intro}, X-ray spectral measurements reveal
substantial optical depths to photoabsorption, $\Nh\sim
10^{22}$~cm$^{-2}$, in the directions of GRBs. Such column densities 
are similar to those of giant molecular clouds in the Milky Way and
therefore imply number densities $n\sim 3\times
10^2\,\mathrm{cm}^{-3}(\Nh/10^{22}\,\mathrm{cm}^{-2})(10\,\mathrm{pc}/R)$.
This is about what is needed for the scattered X-ray emission from the
brightest GRBs to be detectable. There is ongoing debate (e.g. Galama
et al. 2001) as to why measured optical extinctions tend to be small even
when the corresponding $\Nh$ is large. A possible explanation is that
the UV and X-ray radiation from the GRB and its afterglow evaporates
dust out to a few times 10~pc (Waxman \& Draine 2000; Fruchter et
al. 2001). We should also mention the issue of ``dark bursts'', 
i.e. GRBs with undetected optical afterglows. It is possible that such
bursts occur in very dense clouds so that their optical afterglow
emission is extinguished (e.g. Reichart \& Yost 2001; Ramirez-Ruiz et
al. 2002).
 
It is interesting to consider here again (see \S\ref{effect}) the case
of GRB~000926, whose afterglow was observed by BeppoSAX and Chandra. From 
analysis of X-ray and optical data, Piro et al. (2001) have inferred that
the jet had an opening angle of $\theta_0\sim 25^\circ$ and expanded
into a dense medium with $n= 4\times 10^4$~cm$^{-3}$ (see,
however, Harrison et al. 2001; Panaitescu \& Kumar 2002). At the time
of the latest Chandra observation ($t\sim 10^6$~s), the afterglow
emission was still easily detectable, with the flux at 0.2--5~keV
being $8\times 10^{-15}$~erg~cm$^{-2}$~s$^{-1}$. GRB~000926 was
discovered by the Interplanetory Network (Hurley et al. 2000), and its
25--100~keV fluence was $6.2\times 10^{-6}$~erg~cm$^{-2}$ (Piro et
al. 2001). Assuming different values for the slope $\gamma$ of the
X-ray part of the GRB spectrum we may estimate that the GRB~000926
fluence at 0.3--5~keV was $4\times 10^{-7}$, $2\times
10^{-6}$~erg~cm$^{-2}$ and $1.2\times 10^{-5}$~erg~cm$^{-2}$ for
$\gamma=-1$, $-1.5$ and $-2$, respectively. We can then estimate from 
equation (\ref{flux}) the scattered X-ray flux during the last Chandra
observation (assuming that $t\ll t_1$ so that $f(t)\sim$~a few):
$\flux\sim 10^{-16}(1+z)^{-1}$, $4\times 10^{-15}(1+z)^{-1}$ and $2\times
10^{-14}(1+z)^{-1}$~erg~cm$^{-2}$~s$^{-1}$ for the $\gamma$ values
given above. Therefore, if the density of the medium around GRB~000926
is indeed as high as implied by the analysis of Piro et al. (2001),
then the scattered flux already could have constituted a significant fraction
of the observed X-ray flux two weeks after the burst. In this case,
however, the size of the scattering cloud must not be too large
($R\lesssim 10$~pc). Otherwise, it would be difficult to explain the
marginally detected column density $\Nh\approx 4\times
10^{21}$~cm$^{-2}$ (Piro et al. 2001), even taking into account the
photoionization effect of the GRB on the ambient medium
(B\"{o}ttcher et al. 1999).

%%%%%%%%%%%%%%%%%%%%%%%%%%%%%%%%%%%%%%%%%%%%%%%%%%%%%%%%%%%%%%%%%%%
\section{Conclusions}
\label{conclude}
%%%%%%%%%%%%%%%%%%%%%%%%%%%%%%%%%%%%%%%%%%%%%%%%%%%%%%%%%%%%%%%%%%%

We have described a new observational method that enables us to
determine the typical ambient medium density of bright GRBs as well as their
degree of collimation in an almost model-independent way. This
method can hopefully be employed with the existing Chandra and
XMM-Newton X-ray telescopes.

A suitable observational strategy would be to schedule an observation
of the site of a very bright ($\flu\sim10^{-5}$~erg~cm$^{-2}$) GRB
a few months after the burst. Should some X-ray flux be detected
during this observation, an additional one or two observations should be
carried out several months or years later so that the light curve of
the scattered emission could be roughly reconstructed and the GRB 
opening angle could be estimated [from equation (\ref{theta0})]. If the
first observation fails to yield a significant flux, an interesting
upper limit, $\sim 10^3$~cm$^{-3}$, on the density of the medium
surrounding the GRB on parsec scales can be derived [from equation
(\ref{dens})]. It will take a  $10^5$~s observation with Chandra or
XMM-Newton to detect a scattered flux of a few $\times
10^{-16}$~erg~cm$^{-2}$~s$^{-1}$ at 0.3--5~keV. We emphasize that it
is sufficient to collect a total of several photons from the GRB
direction, as no detailed spectral information is needed. We also note
that a flux of $5\times 10^{-16}$~erg~cm$^{-2}$~s$^{-1}$ corresponds
to an equivalent isotropic luminosity of order $10^{42}$~erg~s$^{-1}$ for a
burst at $z\sim 1$. Therefore, the scattered GRB emission should
outshine the X-ray emission of any non-active host galaxy.

Since GRBs with $\flu\sim10^{-5}$~erg~cm$^{-2}$ are very rare events
(at best a few per year), constant monitoring of the whole sky with an
instrument sensitive to X-rays is crucial. After the end of the CGRO
mission, such a capability will be provided by the Swift
satellite\footnote{http://swift.gsfc.nasa.gov/}.

\acknowledgements
We thank Eugene Churazov and Rodion Burenin for helpful comments. SS
acknowledges support from a Peter Gruber Foundation Fellowship. This
research was partly supported by the Russian Foundation for Basic
Research (projects 00-02-16681 and 00-15-96649) and by the program of
the Russian Academy of Sciences "Astronomy (Nonstationary
astronomical objects)".

\end{document}